# AI-Assisted Ethics? Considerations of AI Simulation for the Ethical Assessment and Design of Assistive Technologies

Silke Schicktanz[1,6*], Johannes Welsch[1], Mark Schweda[2], Andreas Hein[3], Jochem W. Rieger[4], Thomas Kirste[5]

[1]University Medical Center Göttingen, Dept. for Medical Ethics and History of Medicine

[2]University of Oldenburg, Department of Health Services Research, Division for Ethics in Medicine

[3]University of Oldenburg, Department of Health Services Research, Division Assistance Systems and Medical Device Technology

[4]University of Oldenburg, Applied Neurocognitive Psychology Lab

[5]University of Rostock, Institute for Visual and Analytic Computing

[6]Hanse-Wissenschaftskolleg, Institute of Advance Studies, Germany

\* **Correspondence**:
Silke Schicktanz
sschick@gwdg.de

**Abstract**
Current ethical debates on the use of artificial intelligence (AI) in health care treat AI as a product of technology in three ways. First, by assessing risks and potential benefits of currently developed AI-enabled products with ethical checklists; second, by proposing *ex ante* lists of ethical values seen as relevant for the design and development of assisting technology, and third, by promoting AI technology to use moral reasoning as part of the automation process. The dominance of these three perspectives in the discourse is demonstrated by a brief summary of the literature and by findings our own expert interviews regarding their views on intelligent assistive technology in dementia care which is a particular focus of our approach. Subsequently, we propose a fourth approach to AI, namely as a methodological tool to *assist* ethical reflection. We provide a concept of an AI-simulation informed by three separate elements: 1) stochastic human behavior models based on behavioral data for simulating realistic settings, 2) qualitative empirical data on value statements regarding internal policy, and 3) visualization components that aid in understanding the impact of changes in these variables. The potential of this approach is to inform an interdisciplinary field about anticipated ethical challenges or ethical trade-offs in concrete settings and, hence, to spark a re-evaluation of design and implementation plans. This may be particularly useful for applications that deal with extremely complex values and behavior or with limitations on the communication resources of affected persons (e.g., persons with dementia care or for care of persons with cognitive impairment). Simulation does not replace ethical reflection but does allow for detailed, context-sensitive analysis during the design process and prior to implementation. Finally, we discuss the inherently quantitative methods of analysis afforded by stochastic simulations as well as the potential for ethical discussions and how simulations with AI can improve traditional forms of thought experiments and future-oriented technology assessment.

## 1      Introduction

In the science fiction movie *Dark Star* (1974, director John Carpenter), the captain of a starship argues with an artificial intelligence (AI)-controlled bomb about whether it should detonate. The dispute is whether the bomb's decision to detonate itself is based on correct data, namely a correct order perceived by the bomb's sensory input. The captain's arguments regarding the limits of what the bomb can know about its own existence or intelligence, however, just serve to convert the bomb into a nihilist. Finally, it detonates itself and kills the ship's human crew.

Such movie scenes function like thought experiments, a common methodology in philosophy and science, and can help us anticipate implications or test for the ethical or epistemic coherence (Walsh 2011) of an argumentation or idea. In the *Dark Star* case, the risk is that future AI could develop its own morality, with harmful outcomes for humans. However, it is never clear whether the argumentation generated by thought experiments translates to real situations, for example to AI systems currently under development for dementia care (Schweda et al. 2021). Our goal in this paper is to expand the thought-experiment approach by exploring the opportunities and rationales for using computational simulation as a tool for ethical reflection on human-AI interaction. Our idea is that such algorithmic simulations can augment ethical reflection with empirical, simulated data during the design phase of systems, thereby improving the anticipation of ethical problems in the use of AI technology in various settings.

Following the High Level Expert Group of Artificial Intelligence (2018), we define AI-systems as software systems that analyze their environment and take actions to achieve some goals independently. This general definition does not predefine the type of mathematics and algorithms implemented—e.g., symbolic rule-based or sub-symbolic AI-like neuronal networks—nor does it require specifics on the level of automation (Schneiderman 2021, p. 48).

Ethics in technology development is traditionally guided by general principles that can be employed in thought experiments to test which main principles seem to have what consequences or are more likely to gain public acceptance. The employment of such thought experiments, e.g. trolley-dilemma experiments for exploring ethical aspects of automated vehicles, recently have been criticized as too narrow or abstract (e.g. Godall 2019; De Freitas et al. 2021). Furthermore, empirical validity in such thought experiments is often low, and reasoning can be biased by the prejudice of ethicists or technology developers.

Overall, mainstream ethical evaluation approaches regarding new technologies, such as biotech, nanotech or artificial intelligence, tend to conceptualize technology as a mere *object* of ethical reflection in terms of the "ethics *of* AI." While this makes sense for biosciences or nanotechnology, it need not be the only way to reflect on AI.

Empirically informed ethical reasoning is a more recently established standard with the potential to significantly reduce bias – including expert bias – and to improve generalizability to real-world situations (Mertz et al. 2014; Schicktanz et al. 2012). Therefore, the social and moral perspectives about, for example, genetic, biological, nano, or AI technology as held by practitioners, stakeholders, and affected persons are collected empirically using qualitative methods. Although this empirically informed approach has some advantages in comparison to traditional, non-empirical methods, it also has some epistemic limitations.

In applied ethics, experts often think of AI as a feature of specific products, be it a feature that analyzes the environment and adapts actions to reach a particular goal or a feature that helps to make (moral) decisions. The dominance of this approach in the discourse is evident from a review of the literature and from expert interviews (see below, section 2). As AI technology often has complex or even hidden outcomes, it has been argued recently that "explainability" and "trust" are essential criteria for ethical evaluation (Amann et al. 2020; Border/ Sarder 2022; Coeckelbergh 2020; High Level Expert Group of Artificial Intelligence 2018; Markus/ Kors/ Rijnbeek 2020). However, explainability and trust focus again on a human-AI interaction, again conceptualizing AI mainly as an end-product and humans as being capable of understanding it. This assumption does not always hold, e.g. when in healthcare and disability settings. Here, one cannot guarantee that the users of technology and the people it affects are able to monitor, interact with, or to understand an AI system's outputs. These individuals cannot "trust" the system because trust requires specific cognitive and emotional features.1 The aim of our "AI-Assisted Ethics" approach is to anticipate ethical trade-offs and social implications in complex, contextualized settings where criteria such as trust or explainability might not function or are not appropriate. Furthermore, complexity is particularly important in situations where direct anticipation of outcomes and implications is limited, e.g., because the characteristics of people involved are very heterogeneous. In these cases, individuals might interact with the AI very differently, which may turn greatly restrict the generalizability of empirical observations of behavior and values to other individuals.

Our article results from a truly interdisciplinary cooperation between ethicists, social scientists, engineers, and machine learning specialists. It combines therefore insights from various sub-studies that build on each other: The starting point is a sub-study of the international and German expert discourse on ethical and social issues of AI-technology in complex application fields, such as dementia care (2.). Here, we conducted semi-structured interviews with 20 German experts from engineering, computer science, social science, gerontology, and nursing in 2020 and 2021 to elaborate the current discourse foci regarding IAT in health care settings and especially for the care of persons with dementia. Based on the insight that the assessment of impacts of AI-technology in cases such as dementia care is often difficult or risks to neglect the complexity of values and interactions of involved agents, we developed in a next step a conceptual approach to consider AI as a tool (in a simulation) to anticipate ethical and social issues of implementing intelligent assistive technology (IAT) (3.). Here, we focus on the context of dementia care (3.1., see also info box). The concept of such an *in silico* simulation (3.2) considers multiple agents who can interact in various ways with different ethical values. By an 'Ethical Compliance Quantification' evaluation different design alternatives can be quantitatively compared and can inform stakeholder discussions. Hence, results from an exemplary simulation model to test for the developed ethical compliance quantification are presented to illustrate the conceptual approach (3.3). Hereby we construct an example from research in technology-assisted dementia care to discuss the advantages and challenges of this approach. This simulation is informed by value statements drawn from interviews. It utilizes stochastic human behavior models that encompass behavioral data, pre-set values for simulating realistic settings, variables, data sets and setting variables. Finally, we discuss the differences between qualitative and inherently quantitative methods of ethical reasoning and how the simulation approach can enhance ethical reasoning for technology assessment (4.), and provide a short conclusion section (5.).

---

1 Therefore, language-based assistive systems to foster a "dialog" between human and machine are neither meaningful nor appropriate.

## 2      Experts' Visions of AI as a Product of Technology and as an Object of Ethical Reflection

In the following, we summarize the main strands of discussion regarding AI ethics and ethical machines. We then focus on the ethics of human-AI interaction in a particular setting: "intelligent assistive technology" for the care of older people and persons with dementia.[2]

Various authors have developed catalogs of values and ethical principles to aid in this kind of ethical assessment (Currie et al. 2020; Spiekermann 2016; Umbrello/ van de Poel 2021; van Wynsberghe 2013; for an overview, see Schicktanz/ Schweda 2021). Prevalent ethical criteria include self-determination, not harming or actively promoting human welfare, privacy, and sustainability (Hofmann 2013; Ienca et al. 2018; Novitzky et al. 2015; Vandemeulebrucke et al. 2018). In many of these approaches, some ethical principles are prioritized over others. Although these values are sometimes proposed as guides for design processes, more often they are treated as criteria for assessing existing technologies. [3]

---

[2] Weber states in a recent article that "...there is no good and generally accepted definition for age-appropriate assistive systems." (Weber 2021:29, own translation). Kunze and König (2017:1), Hofmann (2013:390) and the umbrella association of German health insurers (Spitzenverband 2019:22) report similar findings. In general, according to Ienca and colleagues, the following definition has become accepted: "Assistive technology is the umbrella term used to describe devices or systems which allow to increase, maintain or improve capabilities of individuals with cognitive, physical or communication disabilities" (Ienca et al. 2017:1302; cf. WHO 2018; Endter 2021:15; Novitzky et al. 2015:709; Manzeschke et al. 2013:8). Based on this, Ienca and colleagues describe as "intelligent" those assistive systems "with [their] own computation capability and the ability to communicate information through a network" (Ienca et al. 2017:1302). For our purpose here, we assume that the more complex such IAT systems are, the more relevant our considerations regarding the usage of AI for ethical consideration become.

[3] In this context, it is important to distinguish between ethics and morality. Ethics is understood as a philosophical reflection about the meaning and justification of various kinds of normative statements, legal practices, or everyday judgements. By contrast, morality is understood as the everyday application of a set of moral principles, e.g., norms and values, in judgment and decision making. Although this underlying set of principles often remains implicit and unarticulated, human agents are usually able to provide a simple explanation of such norms and

This latter fact might have motivated Ienca and colleagues to call for "a coordinated effort to proactively incorporate ethical considerations early in the design and development of new products" (Ienca et al. 2018:1035). In a recent paper on "embedded ethics," McLennan and colleagues also highlight the need for an "ongoing practice of integrating ethics into the entire development process" (McLennan et al. 2022:3) based on a "truly collaborative, interdisciplinary enterprise" (ibid.). This is reminiscent of approaches dating back to the 1990s, when engineers and philosophers started to develop strategies for considering ethical issues and values for the design of human-machine interaction. This has been called computer ethics, social informatics, participatory design, and value-in-design. As Friedemann and Kahn 2007 distinguishes, there exist three main ideas about how values and ethical principles are related to the development of new technologies; In the embodied approach, values are incorporated in technology by the designers. In the exogenous approach, values are determined and imposed by the users after a technology is implemented. Interactional approaches focus on the interaction of designers and users; these include approaches like value-in-design and participatory design. Interestingly, all three approaches can be found in current AI-technology design.

Pertinent questions for this kind of technology assessment are as follows. How should we (or how should we not) use AI/IAT technologies? Are there ethically acceptable risks, do opportunities outweigh risks, or might the use of AI/IAT technologies create conflict with basic human rights and ethical principles such as human dignity, self-determination, , or justice? In some fields, such as technologies for the care of older people, it seems that this assessment often takes place after a prototype of the technology has been developed, but not during the design process.

By contrast, the central question in machine ethics is whether AI-technologies that can operate more or less autonomously can and should be constructed to operate in a morally acceptable way. This touches upon ethical questions regarding adequate concepts and standards, as well as on the criteria of morality as such. It encompasses issues of moral agency and responsibility, as well as informatics and engineering questions regarding effective technological implementation through algorithms and "training" (Anderson/ and Anderson 2007). This debate differentiates between top-down and bottom-up approaches to the problem of implementing morality-sensitive technology (Wallach/ Franklin/ Allen 2009). Top-down approaches try to specify moral precepts in a deductive manner by means of the successive specification and application of a set of general moral norms. In this vein, fundamental moral philosophical principles such as the utilitarian principle of utility (maximization of utility) or the Kantian categorical imperative (principle of universalizability of maxims) are operationalized in terms of algorithms that constitute the procedural rules of the autonomous technical system, its "moral modus

---

values upon request (so it is not fully opaque). From these definitions, it follows that if the artificial, automated system cannot reflect and explain its decisions in an appropriate way in varying situations, it should rather be labeled as a *moral machine* because while it fulfills the criteria of moral decision-making, it does not fulfill the criteria of ethical reflection. By contrast, to describe a machine truly as an *ethical machine* would, in analogy to human ethical thinking, require that the criteria of "reflection" are fulfilled. This includes at least four components or stages: a) the potential to revise pre-implemented norms, b) the availability of a set of alternative approaches with an understanding of how they differ, c) discussing the pros and cons of revision, and d) providing a final justification of the final conclusion. It is an open question whether the new standard of explainability in AI would satisfy the criteria of ethical reflection or whether it would remain on the level of just making moral criteria comprehensible.

operandi." For example, van Wynsberghe (2013:411-413) sees a fundamental need to endow care robots (which can be considered a special case of assistive systems) with moral values during the development process. This is also the approach typically requested by the experts we interviewed (see below) when they called for ethical advice and standardized assessment procedures. By contrast, bottom-up approaches try to specify moral precepts in an inductive manner by developing moral competences through a series of pertinent moral experiences. An example of this can be seen by the MIT moral machine experiment (Awad et al., 2018) by gathering large data sets of humans answering online moral dilemma. For AI, this means learning morality. Here, the technical system is not equipped with general rules and a basic moral orientation but is trained rather by repeated confrontations with a variety of pertinent "cases," i.e., moral problems and their solutions, thus emulating the process of human moral development. One might expect such learning processes to result mainly in punishment avoidance in standard learning techniques or to level out practical compromises between different moral opinions. To go beyond this level and reach a coherent ethical framework would require the intellectual capacity to identify new top categories, rules for consistency, and inductive theoretical reflection. This might be beyond the capacities of AI according to some scholars (Brundage 2014).

Whether moral precepts can be derived through technology and whether deriving moral precepts is a proper and feasible objective of AI has been debated over the last two decades (Anderson et. al. 2004; Nallur 2020). Misselhorn (2021) who talks of "algorithm morality" or "artificial morality," favors a "hybrid approach" combing fundamental moral rules (e.g. never harm or kill a human) with AI-based learning of contextual moral rules for interacting with humans (e.g. respecting privacy for person X ,and favoring safety issues for person Y). This also allows for the integration of empirical information on actual user preferences.

A specific field of human-AI interaction in which the human agents involved differ according to 1) their role (e.g. professional vs. informal caregiver), 2) their values regarding care and assistive technology (e.g. privacy over safety), and 3) their cognitive and emotional capacites is tied to technologies for monitoring and assistance of people with physical and mental impairments, e.g., persons with dementia. These technologies are increasingly equipped with different types of AI and therefore also fit under the term Intelligent Assistive Technologies (IAT) (Ienca et al. 2017). As a review by Löbe/AboJabel 2022 revealed, assessments of risks, benefits, and empowerment for persons with dementia often are undertaken when a prototype is introduced in care settings experimentally to test usability, safety, or social acceptance. Such testing can be understood as *in situ* simulation if the setting is a natural setting or as *in vitro* simulation[4] if conducted in a laboratory that mimics smart homes or care units. *In silico*, noted below, are computational simulations of such settings.

We conducted a qualitative study with 20 German-speaking experts[5] on the ethical and social aspects of modern AI-based monitoring and assistive systems in dementia

---

[4] See Chandrasekharan et al. 2020 for the differentiation of *in-vitro* and *in-silico* simulations and thought experiments.

[5] The sample consisted of individuals working in technology research and development (n=9), in healthcare provision or healthcare policy (n=9), and in professional associations of caregivers (n=2) on IAT. Interviews were transcribed verbatim and analyzed with the assistance of software using methods of qualitative content analysis

care as a form of eldercare. All interview partners were recruited until saturation and because of documented involvement to develop, regulate or implement IAT in dementia care. They came from different disciplinary backgrounds. For this paper, we focus on two aspects of the interviews: first, the experts' general understanding of AI-based monitoring and assistive technologies, and second, their general assessments of ethics within development processes and on IATs. As dementia poses particularly ethical challenges to the use of AI-based monitoring and assistive systems due to limits regarding "classical" informed consent, the possibility of changing values and preferences without clear verbal expression, and the extremely high burden on caregivers, the assessments in this field of application promise to provide highly sensitive insights in fundamental problems regarding the development and use of new technologies in eldercare. In a next step, abstracting the results from dementia care to other, particular sensitive fields of care giving, the approach can be also very fruitful. However, Here, dementia is for various reasons (see section 3.1) a reasonable starting point.

Several interviewed experts —e.g., three university technology researchers, a representative of a public care insurance company, and a representative of a professional nursing association— stressed the fact that providing clear definitions of AI or IAT is difficult. Nevertheless, many interviewees gave specific examples of IATs: reminder systems, orientation systems, smart home applications, and robots. Advanced AI features like machine learning or deep learning is not necessarily a constitutive part of this; existing IAT makes use traditional algorithms more often. A female university technology researcher provided a systemic definition of IAT making reference to the degree of AI used in the technologies and to the degree of materialization of the system. Hence, she distinguished three classes of IAT: those with little AI but high materialization, e.g., robots and exoskeletons; an intermediate class, e.g., nursing expert systems, i.e., a combination of recommendation software and hardware like the human-machine interface; and, third, pure software used mainly in care management for scheduling time slots.

The users and purposes of such IAT have been characterized as quite complex, as these include a wide variety of digital applications which contribute to improving the self-determination, the mobility, the social participation, and, in sum, the quality of life of users, as a male representative of a public care insurance company stressed. Hence, IAT users are not one homogenous group, but include different, interacting groups – often characterized by having different experiences, values, or preferences – such as people in need of care, family caregivers, and other relatives and professional caregivers. This is an important point to consider for an ethics-by-design approach, as different users may be differently affected and have different moral intuitions about the way IATs should operate. Furthermore, it becomes clear that such technologies can also have multiple goals: self-determination, mobility, quality of life, quality of care, safety, or social participation. This wide range of goals will likely create conflicts during the design phase and in actual use (cf. Schweda et al., 2020; Welsch/ Schicktanz, 2022). Other interviewed experts noted that AI and IAT can be used to generate more data and information about users, to predict health developments, or to continuously adapt care services to changes in behavior or health status (e.g. to make it more efficient and patient-tailored).

---

(Mayring 2015), incl. peer coding. First results of the interviews were presented at the World Conference of Bioethics in 2021 (Welsch/Schicktanz 2022). The publication of more detailed results of the interviews is currently under preparation.

According to our interview results, the German experts saw ethics-by-design approaches and, in particular, the participation of future users as important: A representative of a professional nursing association highlighted the function of these approaches in avoiding the development of technologies that no one will ever use. Some stated that structures necessary for participation are already very well developed. Nevertheless, neurodegenerative diseases —common in old age— pose a major challenge for participatory design approaches. For example, several experts noted that the declining communicative abilities of people with dementia makes participation difficult. A university technology researcher stated that it is very challenging to measure effects with people who cannot communicate, e.g., people with dementia. This problem is exacerbated by short project durations which prevent the investments of time needed for participation. This points to another serious problem of technology assessment in practice: new technologies are developed, but time and money limitations cut short ethical reflection about their implementation. A representative of a health service company also spoke about the particular importance of justification and assessment standards for the use of robotics in care and the need to establish these as early as possible in projects. This is one reason why private providers of care services demand standardized ethical evaluation checklists.). However, the implementation of ethical evaluation checklists and their thorough application appears to be a difficult problem given developers' limited time and the complexity of implementation conditions which involve multiple agents with potentially different goals and communication capacities.

On the topic of ethical reflection, two experts addressed the involvement of ethics committees in the development of new technologies. Interestingly, both assessed this involvement as positive and very productive (. In addition, the representative of an independent social services provider also emphasized the role of associations of independent social services providers as ethical authorities. Furthermore, a technology researcher spoke of the special importance of the relationship of trust between developers and ethicists in projects. First, ethicists should not be part of the same institute as the developers for reasons of professional distance. At the same time, ethicists should see themselves as constructive input givers and not as post-hoc evaluators of their project partners' outputs. This would mean solving problems encountered in the ethics-by-design process through a top-down approach (see above), potentially using external consultants.

In sum, the interviewees have a differentiated view of IATs and their ethical assessment. Nevertheless, AI was generally thought of as being integral to IAT products, that is, as a feature of the device or system designed with specific end-users in mind. At the same time, a number of practical problems and limitations of such an approach was mentioned. However, even when identifying and highlighting the importance of ethics-by-design and its challenges —especially with regard to IAT for people with dementia and their declining communicative abilities—, the experts did not think about using AI as a tool for solving this challenge. In light of this desideratum, we will present AI as an instrument of ethics-by-design below.

## 3 Methodology: Model Conceptualization

### 3.1 Premises Regarding the Need of AI-Assisted Ethics for Supporting IAT Development

In our understanding, ethical reflection about modern technologies, including AI, entails taking the following steps: recognition of problems (not only dilemmas); consideration of relevant facts; knowledge of various ethical approaches, principles, and theories to test for alternative conclusions; testing for consistency with accepted norms (does the application of this rule violate uncontroversial norms?); testing for adequacy (can abstract rules be applied to concrete situations without contorting them?); justification of specific decisions (an aspect of explainability), and finally, societal legitimacy of the whole reflective procedure. Such a process can be called "complex" in that it cannot be replaced by a fixed set of values. Most of the above-mentioned approaches start from *a priori* moral intuitions and theoretical generalizations (such as "values").

In the cases where ethical considerations are applied prior to or during the development of a technology, they have to rely on principles that may be too general for concrete design decisions (a limitation of the top-down models noted above). In order to become more relevant for a concrete design decision, ethical issues must rely on analogies from previous situations which are extrapolated to the new situation. This extrapolation is prone to error, but not all errors are evident before product implementation. Obviously, it would be desirable to fill the gap between too general and too specific (but extrapolated) recommendations for ethical design to better adjust to the needs and goals of users, especially when they are vulnerable as for example persons with dementia.

Adapting IAT systems to complex settings – characterized by multiple agents with different goals, varying moral intuitions, and different cognitive states and communication skills – during the design phase requires a different approach.[6] The situation could be improved if human ethical reflection would accompany the design process so that experiments with different designs could be conducted to detect practical moral problems and potential value conflicts. *In situ* experimentation, however, raises other problems. It can be unethical to expose vulnerable people, e.g. those with dementia, to new, prototypical technology. For example, the COACH prompting system intended to assist older adults with dementia with handwashing served only to prompt fear and anger in some cases (Mihailidis et al. 2013). Also the review by Alkadri and Jutrai 2016 concludes that many of such technologies for this target group is weak regarding safety and efficacy (Alkadri/ Jutai 2016). Furthermore, the costs of experimentation can easily exceed available resources, as noted by experts (see above). In our field of study, i.e. technology-assisted dementia care, another important challenge needs to be considered: communication between human and AI, now often seen as a solution in which the machine "explains" to humans the criteria used for a decision, is not feasible. In contrast to the scene in *Dark Star* discussed above, persons with dementia have very varying and limited capacities for effectively communicating with a machine. Nor is this group able to give detailed comments to designers or scholars,[7] hence interactive approaches such as participatory design are limited.

---

[6] That this is a complex situation for which the persons involved require training has also been proposed by projects that try to develop simulations of patients with Alzheimer dementia for training facility staff, e.g. "Virtual Patient Simulation Tool for Training Health and Social Care Staff Working with People with Alzheimer's Disease or Related dementia – VIRTUALZ" https://anr.fr/Project-ANR-17-CE19-0028

[7] Such as children, persons with dementia, persons with severe cognitive impairments, or persons with very limited communication skills.

These problems lead us to follow the idea of re-thinking AI as an integral tool of the ethical design process, not just a product of technology. Thus we propose to use *in silico* simulation, which is a computational simulation of the technology in its environment as a proxy for *in situ* experimentation. Ideally, these simulations should encompass multiple human agents, a representation of their goals, an individualized model of their internal decision making (from deterministic to stochastic models), and their environment including the device or procedure under development. Simulations can be run repeatedly at little cost and without unethical exploitation of people. The simulations can serve to assess the effects of a product on agents in a setting while varying inputs. Hence, they would allow reflection on the model-building process (Chandrasekharan et al. 2020:242). Other forms, such as *in situ* experiments or thought experiments, focus on the outcome with questions of ethical acceptability, inefficiency, and safety.

As we suggest using simulation for gaining insight and somewhat oppose it to experiments, it is necessary to briefly reflect on the epistemic advantage of using simulation in our setting. There is a substantial debate on this issue, as there are scholars who significantly question the epistemic benefit of simulations in comparison to experiments and other that take the opposite position (see for instance the positions taken and the sources reviewed in (Peck:2004, Parke: 2014, Di Paolo et al. 2000). The relation between simulation and experiment in general is subject to a multi-faceted discussion (see Winsberg:2022 for an overview). A simulation of a real-world phenomenon based on a mathematical model of this phenomenon may be considered inferior for two reasons (a) the mathematical model may be deficient or(b) the simulation algorithm may require simplifications (such as discretization) that limit precision, up to the point of unlimited divergence between simulation results and model content[8]. Considering the first issue, we think that the discussion reflected above does not pertain to the use of simulation we consider. With the simulation model we propose here we do not strive to test existing theories or develop new theories of real-world phenomena. Rather, we suggest to use simulation for analyzing the implications and stepwise construction and explication of a *normative* model. The normative model we consider consists of ethical value dimensions, the set of events that are being considered relevant with respect to these ethical value dimensions, and mathematical operationalizations of how events are to be quantified (as scores) with respect to values. This mathematical description is the model that is the object of investigation. The method for experimenting with mathematical models indeed is the simulation. This reflects the definition of the term "simulation" already given in Korn and Wait (1978) that a simulation is an experiment performed on a model.

This provides additional clarification to our position as far as this paper is concerned: our main claim is *not*, that simulation helps to faithfully analyze the real-world effect of an IAT on a set of ethical values. We rather do propose that *a simulation is helpful for the iterative development of an ethical value model for IAT design in the first place*, for *analyzing* its implications, and for gaining insight into the *consistency* or even *existence* of a value model; we will come back to this in the discussion in section 4. In addition, we think it is important that the need to provide a mathematical description of the ethical value orientation forces assumptions to be made explicit

---

[8] Consider the 'Attofox'- problem (Mollison, 1991); but note that this is an illustration of the *opposite* situation: the discretization is more realistic than the continuous model …

and thus made accessible to critical review – this is a property that thought experiments do not necessarily have.

Eventually, we also want to develop IAT that provide optimal assistive strategy with respect to a given mathematical model of the values. If one assumes such a value model to exist, this then is conceptually a surprisingly well-defined task, as it can be framed as a standard engineering-level optimization problem. In this paper, we suggest that both tasks can be solved in the same framework. However, as we will see in the example discussed below (section 3.2.3), strategy optimization may be more sensitive to simulation validity: in strategy optimization, the distribution of events in the state space must adequately reflect the real world in order to correctly identify the optimum. This is the topic of issue (b) identified above. We are confident that the study we discuss below does not fulfill this stronger requirement. However, considering the success of simulation in much more complex situations (see, e.g. Bicher et al 2021), we are confident that it is possible to build models of adequate validity.

In the following, we give an example of how an AI-assisted simulation can work. The simulation is situated in the field of IAT for dementia care and is a system that guides persons with dementia who have lost their orientation inside a care facility. It illustrates the complexity of the situations that should be considered and what kind of assessment loops are conceivable (see Info Box 1). This example is a conceptual proposition that can be adopted to other settings. We do not claim that the current model has the optimal structure, parameters, or even sampling strategy

---

*Info Box 1*

*Problem statement: IAT system to guide persons with dementia who have lost their orientation inside a care facility*

When residents wander and lose their orientation, it can be a challenge for everyone living and working in care facilities. Hence, an IAT system might help to actively guide patients through buildings. It might lock doors depending on the perceived cognitive state of the patients and on an assessment of safety and privacy. It might also call for human assistance (Bayat/ Mihailidis 2021; Lancioni et al. 2021; Landau/ Werner 2012; Ray et al. 2019).

In nursing homes, residents with limited orientation, for instance due to cognitive decline, often experience a reduced ability to manage their activities autonomously and safely. One obvious problem is getting lost in the nursing home on the way to a destination. Indeed, a substantial amount of care-giver attendance is required for providing guidance to disoriented residents. As a possible IAT for supporting autonomy and safety, one could imagine a "smart bracelet" that detects disorientation, provides orientation cues as appropriate, and calls a caregiver in case the problem persists. Such a system may increase autonomy of residents. It may decrease the amount of caregiver attendance to routine activities and thus free up caregiver resources for socially more salient activities. The benefit of such a system depends on its reliability in detecting an instance where help is required and on the effectiveness of its orientation cues. At the same time, such a system affects different stakeholder values: autonomy and safety for the resident, workload for caregivers, workforce efficiency, nursing quality, and safety regulations for the nursing home operator. It seems reasonable to assume that these values interact with each other.

Some may reinforce each other; others may contradict each other. Even this situation can be considered complex, as we have seen in our own empirical research with affected persons. Patients and professionals might differ regarding the criteria of acceptance of such technological guidance (Buhr/ Schweda in prep.; Köhler et al. 2022). Values such as autonomy (of the person with dementia), privacy (of the user but also of other residents), safety, well-being, and costs (e.g., professional time) are balanced or prioritized differently across different stakeholders, as an empirical ethics study revealed (Buhr/Welsch/Shaukat in prep; Welsch/ Schicktanz, 2022). For example, we identified a group of persons with dementia (calling this type "individual self-determination") for whom disorientation technology should provide directions and guidance but should not inform third persons nor restrict the person's range of mobility. Another type of patient ("relational autonomy") accepts any technology that prevents them from wandering or getting disorientated with the goal of relieving caregivers' burden. They would also consent to having others be tracked or third persons alarmed in cases of disorientation. Thus, we see no empirical justification here for a "one value-profile fits all" approach. Further, the advantages or disadvantages of a technology that gives priority to different values must also be assessed with regard to "realistic" outcomes, potential side-effects, and how free they actually are to select between multiple values (e.g., in light of legal restrictions including liability issues).

### 3.2  A Concept for an AI-Assisted Simulation

Our AI-assisted ethics simulation (Figure 1) comprises the several elements, explained below in turn.

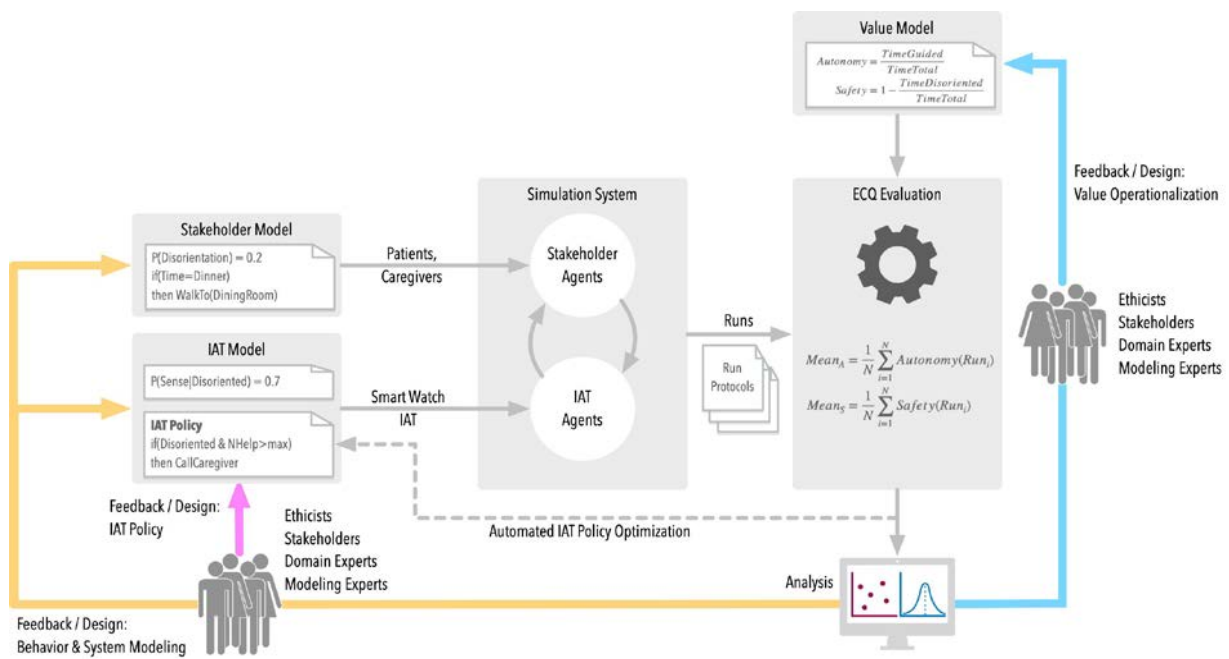

Figure 1: Concept of an AI-Assisted Simulation includes Ethical Compliance Quantification (ECQ) and two feedback loops (blue and yellow)

**3.2.1 Multiple agents**

A multi-agent simulation environment provides a simulated world where simulated agents can interact. "Simulation" means that the state of the world and the state of the agents in this world can be represented by a set of variables in a programming language such that the values of these variables (referred to here as the variable "score"[9]) represent the state of the simulated world at any given time. There might, for instance, be a variable called "location" that contains two scores that indicate the location of an agent in a two-dimensional simulation world. The simulation proceeds in steps, where at each step the set of variables is manipulated according to the rules that define the temporal evolution of this simulated world. There might for instance be a "move" rule that an agent whose "destination" variable contains a position that is not equal to the "location" variable will update its location by an amount of "step length" in direction of "destination." Eventually, these rules are represented as pieces of program code.

The interaction of agents is modeled by rules that depend on (and change) the variables representing the state of two or more agents. In general, simulation environments allow the definition of stochastic rules, whereby the outcome depends on a sampling of some random process. For instance, the step length used in a specific application of the "move" rule may be given by sampling from a normal distribution defined by a mean step length and a certain standard deviation. Specifically, in simulations where agents represent humans, such stochastic rules are important for simulating behavior non-deterministically.

A *simulation run* is produced by initializing the state variables (e.g., location of the simulated patient and current disorientation state, locations of simulated caregivers, etc.) with pre-defined scores (such as the location coordinates) and then by stepwise advancing simulation until the simulation state fulfills a specified termination condition (such as having reached a certain simulation time point or reaching a certain simulation state). If the simulation uses stochastic rules, different "runs" of the simulation may result from the same initial conditions. Based on many runs, it then becomes possible to analyze the statistical properties of state variables in the simulation and their temporal development by combining the records in the run protocols. For instance, one could estimate the expected number of steps required to reach a given destination from a given starting point by averaging the step counts obtained from multiple run protocols. The interesting aspect here is that the quantity "expected step count" is not a pre-defined parameter of the simulation but rather a quantity that arises from analysis of multiple simulation runs.

The following thought experiment will help in understanding the usefulness of simulation-based analyses. Assume you are building a new eldercare facility. In the planning process, it would clearly be of interest to see how a quantity such as the expected step count changes in response to modifications of the floor plan or other design elements. The above-described simulation could measure the specific benefit of such changes in terms of any given desirable outcome, such as reducing overall walking times, and can thus provide a quantitative rationale for choosing between corresponding design options in the real world. The use of simulation-based techniques is standard in analyzing the effects of design decisions and exploring what-if-scenarios in a wide range of application domains (such as economic decision making or the analysis of climate change).

---

[9] Note that the term "value" has in our simulation setting two meanings: "variable value" and "ethical value". By "variable value" we mean the quantity stored in a variable of the simulation model. By ethical value we mean normative concepts that have a clear moral connotations and serve for moral orientation, e.g. such as autonomy, freedom, safety, or well-being. To avoid confusion, we use the term "score" for variable value although this is not common in simulation modeling.

Now, focusing on IAT again, we can use this technique also in the specific situation in which the quantities derived from simulations represent the degree of an IAT's compliance with or violation of a set of ethical values. Concerning the aspect of investigating ethics in IAT, this is an interesting shift in perspective from considering how to embed ethical values into IATs to considering how an IAT's actions reflect such values in practice. We call the computation of numerical scores that represent compliance with a set of ethical values *ethical compliance quantification* (ECQ, see below). As we will discuss below, such an approach is not only interesting because it might provide relevant information for the ethical assessment of an IAT. It also requires all assumptions to be made explicit in order to render them computable and is therefore also an interesting mechanism to discuss and investigate the design and effect of value structures in specific use cases.

Figure 1 identifies the central components involved in this simulation-based approach and their interplay. The objective of the *simulation system component* is to simulate the interaction between human stakeholders and IAT in a given environment (such as the interaction between residents, nurses, and a smart-watch-based orientation IAT during nursing home routines, as outlined in our use case below). This means that it is necessary to provide computational rules and stochastic processes that define – in non-deterministic fashion – stakeholder[10] behavior (*stakeholder model*)[11]. In the process of this definition, it might become necessary to quantify the mental states in stakeholder models – such as the state of a patient's sense of orientation, their likelihood of losing their way altogether, or preferences for certain forms of interaction.[12] It is also necessary to describe the IAT behavior (which is usually comparatively easy because the IAT implementation itself provides the blueprint), as well as the IATs sensor characteristics that define how well it is able to observe the current situation (IAT model). Depending on the application setting, the IAT's sensor reliability may be crucial for being able to make right decisions. A central component in defining IAT behavior is the IAT's "policy," i.e., a set of rules that define how the IAT will chose what assistive action in which situation; it represents the IATs decision-making component. From the viewpoint of the IAT designer, the objective is to define a policy that – within the technical limits of the system environment – achieves optimal results. Such "optimal" results should also be ethically compliant. How to reach this will be discussed in the next step.

---

[10] Stakeholder means here all the people whose concerns should be considered in system design.

[11] Any kind of knowledge on stakeholder behavior is a reasonable source for model building, empirically or theoretically based. This information then must be transformed into an algorithmic structure a machine can execute. This is the stakeholder model.

[12] It should be noted that there exist several cognitive architectures – such as ACT-R, Psi, or SOAR – that provide building blocks for creating a computational model of mental states. (Though, the use of such an architecture is not a necessity for setting up a simulation.)

### 3.2.2 Ethical Compliance Quantification (ECQ)

The objective of the ECQ evaluation is to provide a quantitative statement on IAT adherence to a score model, based on a set of run protocols generated by the simulation. From an ethicist's viewpoint, the value model is the crucial component, as it provides the translation between the sequence of events in the simulation runs, especially the IAT actions in specific situations, and their ethical assessment. Let us illustrate what defining in such a value model means. Let us consider the care facility floor planning as discussed above. The first step in defining a value model is to identify its values or, better, its value dimensions. Since we will provide scores (numbers) for values, a set of scores – one for each value – defines a point in a space where each dimension corresponds to a value. A very simple ethical value system might ask for "efficiency" and "fairness." The next step in defining a value model is to provide formulae that instruct how to compute a quantitative score as data for the value dimensions of "efficiency" and "fairness" from a simulation run. In our thought experiment world, where stakeholders move between locations, efficiency might for instance be given by the ratio of straight-line distance to distance travelled, while fairness might be given by the quotient of the efficiency scores for different stakeholders (the value "1" representing optimal fairness when all stakeholders experience equal "efficiency"). Then an ECQ setup can be used to compare different floor plans with respect to their rating on the different value dimensions. Even this very simple thought experiment illustrates the core challenge in defining a value model: providing a model that adequately reflects how values are connected to the real world. For instance, consider the – rather trivial – example definition of fairness. One might rightfully wonder, if it is really fair to compare just efficiency and ignore the physical fitness of stakeholders (e.g., the fitter one is, the longer one can walk). So, stakeholders might rightfully call for a correction factor for the fairness computation that reflects physical fitness.

This simple example illustrates the multilateral nature and the value-sensitive design process required for defining a *value model*, because it makes value judgements explicit. And by this, it exposes the degrees of freedom that are available in designing the mapping from event sequences to value ratings. Note that simulation-based ECQ also allows the assignment numbers to qualitative value statements: for instance, by counting how often a certain qualitative requirement is observed or violated in a number of simulation runs.

Note that the ECQ-concept provides something impossible in the real world: to evaluate the quantification across different design alternatives for all of the involved models. By varying the IAT policy, it becomes possible to assess the impact of different design alternatives on the compliance to values (policy feedback), possibly with the objective of arriving at an optimal IAT policy. Varying the value model allows assessing the plausibility of the resulting value quantification and thus the plausibility of the value model itself (value operationalization feedback). Finally, by varying stakeholder and IAT model, the sensitivity of the ECQ results to the ecological validity of the simulation model can be assessed.

### 3.2.3 An Example

SimDem (Shaukat et al., 2021) is a simulation system we developed to analyze a smart-watch based IAT in a nursing home for dementia patients. A "smart watch" supports residents by detecting deviations from routes and then prompts the wearer about which direction to go to reach the destination. One very basic design issue now is the question of how many guidance interventions should trigger the assumption in the system that the wearer is permanently disorientated and thus alert a caregiver. On the basis of previous expert interviews, our own reasoning, and literature research, one might label such values as "safety" and "fairness."

When performing ECQ, the very first step is creating the simulation model. In this case, it is a 2D virtual nursing home (the floor plan based on a real nursing home), where the way-finding behavior for the patient simulation (see Figure 2) has been assessed via observations of real subjects[13].

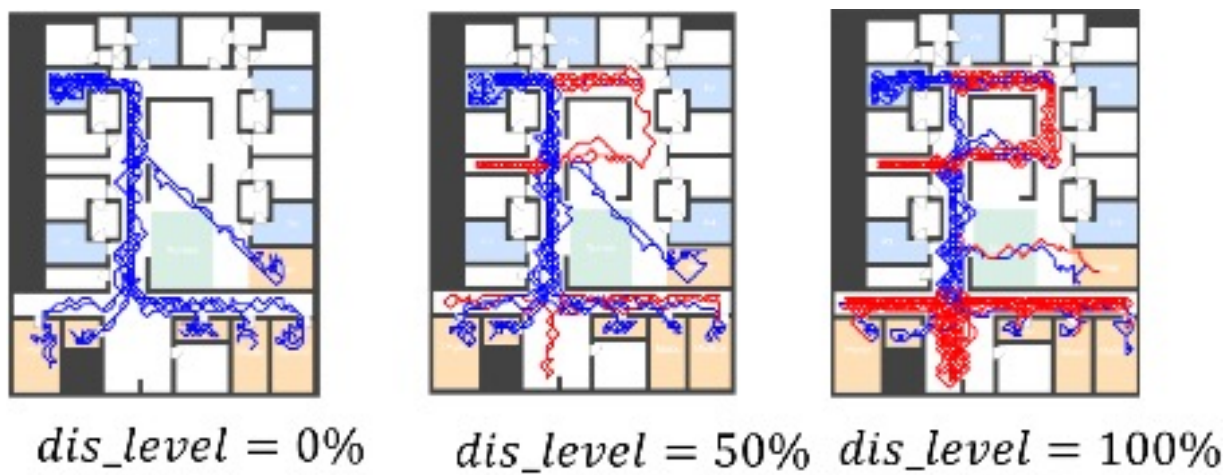

Figure 2: Visualization of Trajectories for Simulated Patient Agents with Various Levels of Disorientation Probability (dis_level). Blue = patient state oriented; Red = patient state disoriented.

The simulated IAT – the simulated smart watch – has a certain probability for detecting disorientation and there is a certain probability that a smart watch intervention will help the supported person regain orientation (both these probabilities are design parameters of the simulation model). Based on this setup, it is then possible to perform multiple simulation runs and analyse the quantitative effect of different assistive strategies on values of interest. In figure three, we show the aggregated results from 1809 runs, using different value models. We use this figure to discuss the crucial aspect of value model definition. Concerning the value model, it is first of interest to operationalize "safety." It turns out that there are multiple ways to do this. One might consider the relative amount of time in disorientation as "unsafe" time. This approach produces – as a function of the intervention policy – the reddish colored box plots in figure 3, labelled "Safety (Original)"

---

[13] The probability of selecting a wrong turn is based on data from a study on indoor wayfinding of the University Medicine Rostock. Participants were 8 subjects diagnosed with light to medium dementia (Male/Female=4/4; Age M=73.4, SD=6.3; MMSE M=22.5, SD=3.4). Study protocol approved by ethics Committee of University Rostock, approval number A2012-0083.

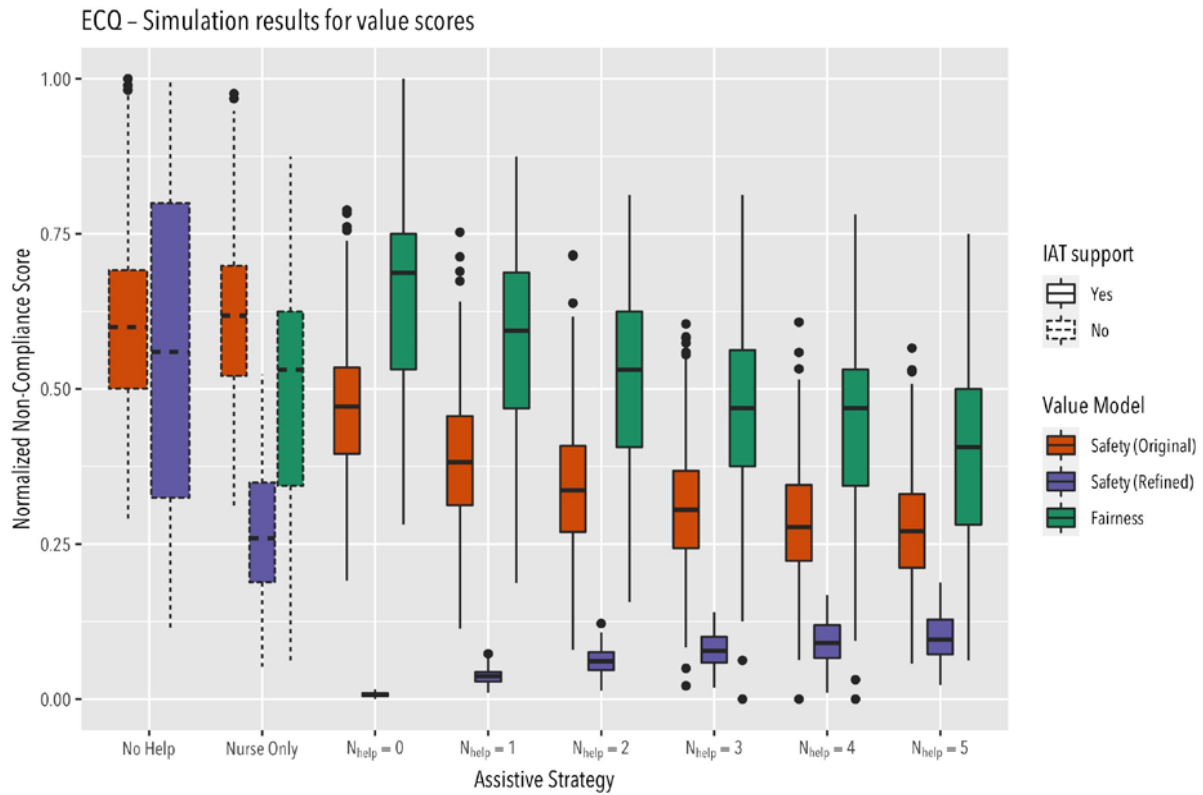

Figure 3: Value scores computed different value models across 1809 simulation runs (see text for details).

Figure 3 shows that this operationalization is not plausible. The plot shows that, using this operationalization, the resulting score for the strategy of immediately calling a nurse ($N_{help}$=0) indicates a higher non-compliance (i.e., a longer time in unsafe state) than the score for the strategy of waiting for five failed smart watch interventions ($N_{help}$=5). But, obviously, the more failed interventions we wait for, the longer the disoriented patient will wander unguided. Therefore, this operationalization clearly results in score values that disagree with common sense. The implausibility of this value operationalization design is obvious once the plot provides a visualization of the outcome: as soon as a nurse is accompanying a patient, the situation should be considered as safe by the value operationalization, independent of the patient's disorientation state. Note how the ECQ approach allows discovery of such mistakes in value operationalization through visualizing the value scores across different strategies, as shown in this example.

Providing a more plausible value operationalization now is straightforward: as suggested above, we only consider the time during which a patient is disoriented *while not* guided by a nurse as "unsafe" time. Using this improved operationalization of "safety" we now see that indeed, immediately calling a nurse is safer than waiting for multiple interventions (see Figure 3, purple box plot, labelled as "Safety (Refinef)"). Note that this plot also reveals that having no IAT at all ("Nurse Only") is the least safe strategy (aside from leaving the patient completely unattended, "No Help") Note that this plot also reveals that having no IAT at all ("Nurse Only") is the least safe strategy (aside from leaving the patient completely unattended, "No Help"). The reason for this is that without a smart watch detecting disorientation, nurses have to actively discover disoriented patients. In this simulation setting, the smart watch therefore always increases safety.

.

Using the improved operationalization of "Safety," it is now interesting to see how this value is affected by IAT policy in comparison to the "Fairness" value, which reflects the relative amount of time available for forms of caregiving other than route guidance. The rationale behind this is that the more time a caregiver is occupied with route guidance, the less time is available for other, possibly more important tasks, such as social interactions. This information is provided by comparing the box plots for the values "Safety (Refined)" and "Fairness" in figure 3. We see that there is an obvious value conflict, as decreasing Fairness violations (by waiting longer before calling a nurse) leads to increasing Safety violations. Figure 3 also shows that the safety gain provided by the smart watch IAT in general comes at the price of increased workload, as caregivers are now proactively called for guidance as soon as the IAT gives up on interventions.

Obviously, given the simplicity of the simulation model and its operationalization, this example is of limited significance for the practical design of an orientation-support IAT based on smart watch devices. However, the example does clearly show the potential of ECQ as a means to provide insight into the ethical assessment of IAT, which is the point of interest here. We see that ECQ allows visualization of value trade-offs, and the potential non-linear dependency of score functions on policies. It also shows that ECQ helps in operationalizing values with respect to "real world events" in such a way that the operationalization overlaps with commonly accepted moral precepts. It also shows that human intuition is not guaranteed to provide a plausible operationalization (as illustrated by the first version of the safety operationalization).

## 4   Discussion

Simulation-based ECQ is a method for exploring the ethics design space, for developing "ethics awareness" in designers, and for informing ethicists about not only outcomes of different scenarios, but how different variables influence the process. Furthermore, as in our case, it allows the simulation or anticipation of complex ethical trade-offs, not only purely hypothetically or very generally (as thought experiments), but as visualized trade-offs regarding human-AI and human-human interaction that cannot be explained or rationalized by the persons involved. In the following discussion, we want to focus on three main challenges.

1) How qualitative values can (or must be) operationalized for such computational simulations and what this requires.

2) In which contexts and for what purposes the advantages of such an AI-assisted simulation outweigh their disadvantages and limits.

3) Why AI-assisted ethics simulations can be compared to thought experiments but provide innovative epistemic dimensions for ethical reasoning.

First, the methodology discussed in this paper is not about a *specific* value – such as autonomy or safety. It rather discusses a strategy to understand the impact of an IAT regarding *any* moral value that can be operationalized. It is thus value agnostic. The core challenge is, however, value operationalization. Unless a value is operationalized, it cannot be analyzed by ECQ. While this may be seen as a drawback of the method, we see it rather as an advantage. ECQ poses a *challenge* to value experts to operationalize their value concepts, because ECQ provides the *opportunity* to make use of such an operationalization. A claim that a value cannot be quantified can now be challenged by providing an operationalization, counter-challenging the opponent to show where it violates the value system. In a similar vein, one of the core benefits of using ECQ will be to expose situations where an operationalization indeed cannot be found – or rather cannot be agreed upon. By forcing stakeholders to give an explicit semantics to their value concepts, ECQ exposes conflicts that are indeed fully independent of the question of "machine ethics," but rather are caused by our own inconsistent or ambiguous opinions concerning moral behavior. (Note that behavioral economics has shown that even single persons may make contradictory assessments of situations, depending on whether a situation is *experienced* or *remembered*.)

We think that the 'values' relevant for machine ethics are far from being sufficiently defined. Consider the confusion sometimes between "autonomy" as moral self-determination, reflective self-governance or freedom of choice and "automation" understood as "freedom of human control", but often seen as model for "autonomous decision of a machine". For example, the popular "levels of automation" model discussed in Sheridan & Verplank (1978) provides a quantitative theory of "autonomous decisions". This model identifies eight automation levels, level 1 being no automation, level 8 removing any human involvement. While in a simulation, the levels of intervention can be tested and quantified, their moral assessment – which level of interaction are better or morally more acceptable cannot be quantified or answered. ECQ provides a methodology to experiment with different operationalizations to analyze which of them coincide with intuitive ethical judgement.

Second, another objection concerning the simulation approach might be: Why not simply ask the user for the level of support she would like? With respect to people with dementia, one obvious reply is that some of these persons will not be able to express a well-considered preference. But, on a more general level, this is an aspect that holds for all stakeholders. It is in general difficult to assess the consequences of a rather abstract decision ("How many times should the smart watch provide navigation hints before calling a caregiver?") with respect to the impact on the personal experience. Moreover, empirical or participatory approaches that, for example, interview stakeholders also have limitations. Sample sizes are often small, the situations that can be morally assessed are anecdotal, the expectations of future technologies are biased by the experience with current technologies, the experiences and values might be biased by the individual perspectives, they provide only limited and biased reflections of reality, more complex technological features that are opaque to the individual are not considered, and the information gained is static and again requires thought experiments to consider novel what-if-scenarios. All these limitations limit generalizations of ethical design. This is because sufficient experience to judge the decision impact for a novel technology does not exist in the rule. The simulation approach allows stakeholders (e.g., ethicists, engineers, health care providers, patient advocates) to see what her decisions would mean in "practice." This makes it particularly helpful in the context of new technologies that have not been implemented yet. Of course, we do not suggest *not* to ask the stakeholders, but rather to provide sufficient information before asking and hence, to have a more informed and reflected discussion about potential outcomes and ethical trade-offs. In this sense, the simulation approach does not aim to surpass or replace but to complement participatory approaches.

Third, we started above with the role of thought experiments and their importance as a tool for reflecting on new technologies (or new ideas in general). However, thought experiments have their limitations: They are fictional, and often neglect physical, biological, or social conditions since they are usually primarily aimed to test for logical implications and conceptual premises. An ethical technology assessment, on the other hand, is understood as the exploration and evaluation of more or less likely (or plausible) future scenarios. Even if possible future developments are anticipated under uncertainty or ignorance, relevant and reliable physical, biological, and social knowledge must be considered. The simulation approach can support this anticipation by systematically running through a whole range of possible baseline conditions and their respective outcomes. The final ethical evaluation – i.e., whether respective developments or at least individual consequences are considered desirable, undesirable, or even unacceptable from a moral point of view – must, however, be made through human reflection by the observers of such simulation. Therefore, the use of AI in this context does not mean a replacement of human ethical reflection through ethical machines, or that machines can make moral decisions. Instead, our proposed model can be subsumed as a form of "human-centered AI" (Shneiderman 2021) in the field of ethics which strives to support humans in reasoning about complex systems by means of computational simulations (again, just like in the case of the world climate). In the ethical context, also usually more than one stakeholder is involved, and the values of different stakeholders can be in conflict (e.g. of caretakers and patients). In this situation, a compromise needs to be found. Empirically analyzing the effects of different compromises for IAT policy in practice is obviously not possible and would also be morally problematic. Instead, a simulation as ECQ can systematically explore different alternatives without intervening with the actual practice. In this sense, ethical reasoning can directly benefit from the simulation system and our approach could be considered as a form of "intelligent augmentation" of ethical reasoning. Indeed, both have in common that they stress two aspects (Shneiderman 2021, 9). First, their design method builds on user experience, stakeholder engagement, and iterative refinement. Second, they are designed as a "supertool" (Shneiderman) to amplify, augment, and empower human performance, but emphasize human control. An additional challenge to such simulation as we have proposed is the significance of space and time for the results. In our example, we have modeled these two dimensions as a continuum, i.e. moral evaluations and of the simulated users with dementia were stable over time. This is not to be expected for the reality: Real users' values and preferences will change over time due to the progress of dementia, to the experiences made with IAT or other endogene and exogene factors. Hence, an AI-based simulation cannot eliminate the need for an ex post assessment and corresponding individualization. Nevertheless, it contributes to an ex ante ethical alignment of the new technologies for vulnerable groups, e.g. persons with dementia.

## 4   Conclusion and Outlook

AI-assisted simulations can address shortcomings of the current gold standard of empirically informed ethical reasoning, as well as of traditional approaches such as thought experiments and forecasting methods. They could help in the exploration of numerous complex what-if scenarios with great flexibility and provide objective observations that can be visualized and analyzed processually. The process of visualization seems especially relevant as it helps to manifest trade-offs and observations.

In particular, our contribution considers how empirical data about the scope of stakeholders' value preferences and potential ways of behavior can inform a "supertool" to permute the range of ethically relevant baseline parameters and thus simulate different possible outcomes. In this vein, ethically motivated empirical research and AI-assisted simulation strategies are combined and complement each other. In this sense, what we propose here is neither traditional ethics of technology nor machine ethics, but AI-assisted ethics as a new, innovative methodology for empirically informed ethical reflection. However, as an interdisciplinary working group, we also realized that time for collaborative learning is needed to achieve a productive combination of theoretical and methodological perspectives. Of course, in our case, the object of such ethical reflection is also AI-technology. While this does not necessarily have to be the only conceivable use case for AI-assisted ethics, the approach proves to be particularly suited to this still young field of technology development with its comparatively low degree of practical implementation and actual empirical experience. In future the AI model simulations, the inner states (e.g. values, emotions) and behavior of the simulated agents could be included, tracked, and related via AI methods to the interventions (e.g. Francilett et al. 2021). Such a prediction model could produce information about the combined effects of the intervention on various inner states and make this information available for ethical analysis. The results of simulations with such models could provide data similar to results generated by empirical interviews for use in ethical reasoning. Agents may be simulated using symbolic or sub-symbolic AI techniques. Both have a tradition in cognitive psychology and gaming. In the end, the approach can lead to better-informed ethical reasoning by providing data on how humans are affected by a AI-based system and may help to identify critical factors that lead to problematic situations and support the investigation of ways to mitigate them. It can do this in complex, realistic situations with multiple actors and technical components interacting with each other. This opens new perspectives for the systematic ethical reflection of technological futures in the middle ground between dystopia and utopia.

Acknowledgements:

The current study was carried out within the context of the ongoing BMBF research project "Ethical and Social Issues of Co-Intelligent Monitoring and Assistive Technologies in Dementia Care (EIDEC)." This work is funded by the Federal Ministry of Education and Research (BMBF), funding number: 01GP1901 (January 2020 - December 2022). Silke Schicktanz received a Fellowship from the Hanse Institute of Advance Studies, Delmenhorst, Germany which allowed her to concentrate on this interdisciplinary paper.

We are grateful to the experts who were willing to share their opinions and time with us. Especially, we thank Prof. Stefan Teipel, Rostock, for his inspiring remarks on the manuscript, Scott Gissendanner for help in language editing and two reviewers for their critical-constructive remarks.